\documentclass[10pt,aps,prb,twocolumn,amsmath,amssymb,floatfix,superscriptaddress]{revtex4-2}
\usepackage{graphicx}
\usepackage{bm}
\usepackage{bbm}
\usepackage{siunitx}
\usepackage[
colorlinks,
citecolor=blue,
linkcolor=blue,
urlcolor=blue,plainpages=false,pdfpagelabels
]{hyperref}
\usepackage[capitalize]{cleveref}
\usepackage{braket}

\renewcommand{\v}[1]{\ensuremath{\mathbf{#1}}}
\newcommand{\hc}{\text{H.c.}}

\newcommand{\tvk}{\delta{\v{k}}}
\newcommand{\tk}{\delta k}
\newcommand{\tq}{{\delta q}}
\newcommand{\NSN}{\text{NSN}}
\newcommand{\DeltaP}{\Delta_{\text{p}}}

\newcommand{\vkappa}{{\boldsymbol{\kappa}}}

\begin{document}

\title{
Supercurrent-enabled Andreev reflection in a chiral quantum Hall edge state
}

\newcommand{\affB}{SUPA, School of Physics and Astronomy, University of St Andrews, North Haugh, St Andrews KY16 9SS, United Kingdom}
\newcommand{\affA}{Department of Physics and Materials Science, University of Luxembourg, L-1511 Luxembourg, Luxembourg}
\newcommand{\affC}{Institut f{\"u}r Mathematische Physik, Technische Universit{\"a}t Braunschweig, 38106 Braunschweig, Germany}
\newcommand{\affD}{Laboratory for Emerging Nanometrology, 38106 Braunschweig, Germany}

\author{Andreas Bock Michelsen}
\affiliation{\affB}
\affiliation{\affA}

\author{Patrik Recher}
\affiliation{\affC}
\affiliation{\affD}

\author{Bernd Braunecker}
\affiliation{\affB}

\author{Thomas L. Schmidt}
\affiliation{\affA}

\date{\today}

\begin{abstract}
A chiral quantum Hall (QH) edge state placed in proximity to an $s$-wave superconductor experiences induced superconducting correlations. Recent experiments have observed the effect of proximity coupling in QH edge states through signatures of the mediating process of Andreev reflection. We present the microscopic theory behind this effect by modeling the system with a many-body Hamiltonian, consisting of an $s$-wave superconductor, subject to spin-orbit coupling and a magnetic field, which is coupled by electron tunneling to an integer QH edge state. By integrating out the superconductor we obtain an effective pairing Hamiltonian in the QH edge state. We clarify the qualitative appearance of nonlocal superconducting correlations in a chiral edge state and analytically predict the suppression of electron-hole conversion at low energies (Pauli blocking) and negative resistance as experimental signatures of Andreev reflection in this setup. In particular, we show how two surface phenomena of the superconductor, namely Rashba spin-orbit coupling and a supercurrent due to the Meissner effect, are essential for the Andreev reflection. Our work provides a promising pathway to the realization of Majorana zero modes and their parafermionic generalizations.
\end{abstract}

\maketitle

\section{Introduction}

When electrons are confined to a two-dimensional interface, thus forming a two-dimensional electron gas (2DEG), a strong externally applied magnetic field causes the quantum Hall (QH) effect, where the bulk becomes an insulator while the edges host chiral one-dimensional (1D) conducting states. Some features of the quantum Hall effect can be understood in a semiclassical picture: Electrons can form cyclotron orbits in the two-dimensional bulk but near the 1D edge the orbits are disturbed and turn into ``skipping orbits'' along the edge, effectively leading to chiral charge transport.

\begin{figure}[h!]
\centering
\includegraphics[width=0.8\columnwidth]{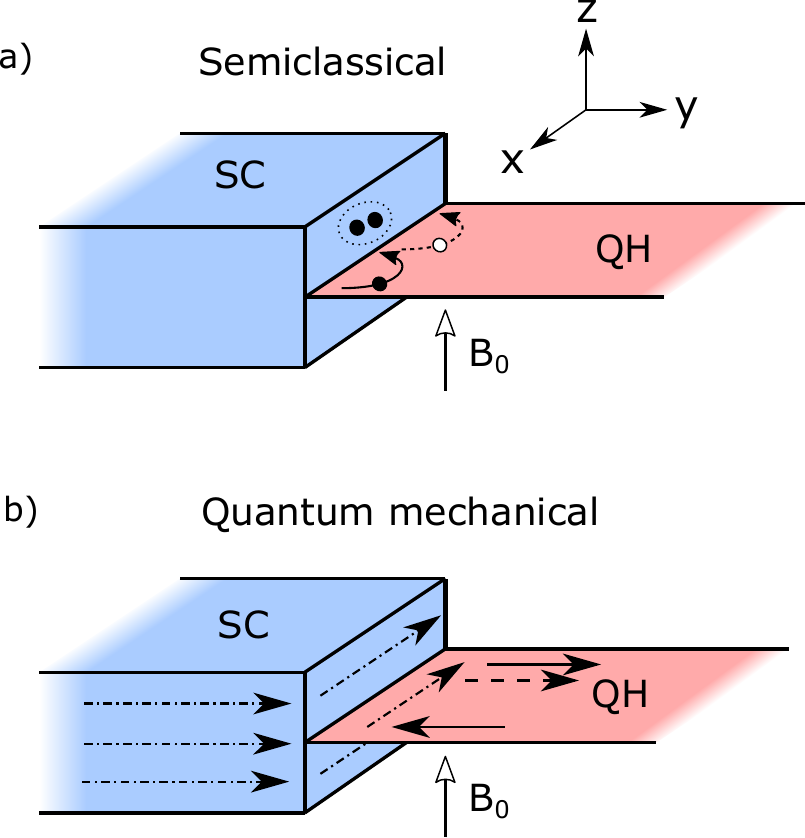}
\caption{In the semiclassical picture (a), an electron performs a skipping orbit along the QH edge, changing from electron to hole with each Andreev reflection at the QH/SC interface. As the hole is retroreflected with opposite momentum of the electron, no momentum transfer occurs. In the quantum mechanical view (b), an incoming electron (solid arrows) can tunnel across the QH/SC interface, which induces particle-hole mixing (dash-dotted arrows) in the chiral QH edge state. This can result in an incoming electron being transmitted as a hole (dashed arrows) across the interface. For a clean interface, tunneling conserves momentum, and can only occur in the presence of a supercurrent at the SC surface. }
\label{fig:sketch}
\end{figure}

When such an edge state is coupled to a superconductor (SC), superconducting correlations can be induced in the former through tunneling of electrons across the interface. However, as single-particle tunneling into a superconductor is strongly suppressed due to the energy gap, the dominating tunneling process will involve pairs of particles performing Andreev reflection. This can be understood as two electrons from the edge state tunneling as a pair into the superconducting condensate or, in the semiclassical picture, as an edge electron being retroreflected as a hole, leading to an effective hybridization of the QH edge state and SC surface \cite{takagakiTransport1998,hoppeAndreev2000,giazottoAndreev2005,khaymovichAndreev2010}; see \cref{fig:sketch}(a).

Andreev reflection in the absence of a strong magnetic field has been well studied in both experiment and theory for decades \cite{klapwijkProximity2004}. The coexistence of superconductivity and a QH phase, however, poses a challenge because the magnetic field must be strong enough to reach the QH phase of the 2DEG but not so strong as to suppress superconductivity. This is why experimental indications of Andreev reflection in a QH edge were observed only very recently through transparent interfaces with high-field superconductors \cite{wanInduced2015a,caladoBallistic2015,benshalomQuantum2016}. Signatures of Andreev reflection have been observed in the form of Cooper pair transport between two superconductors through a graphene sheet in the QH phase \cite{ametSupercurrent2016,sahuQuantized2021}, as well as signatures of electrons being converted into holes in current passing a superconducting electrode \cite{zhaoInterference2020,hatefipourInduced2021}. The details of these signatures have been found to depend on disorder in both the 2DEG \cite{manescoMechanisms2021} and the SC \cite{kurilovichDisorder2022}. Crossed Andreev reflection, where an electron is transmitted across a narrow SC as a hole, has also been observed in graphene in both the integer \cite{leeInducing2017} and fractional \cite{gulInduced2020} QH phases, and is under active theoretical investigation \cite{schillerInterplay2022,galambosCrossed2022}.

In addition to the demonstration of a fascinating effect, these experiments pave the way for novel applications in quantum technology. If two counter-propagating edge states are coupled via crossed Andreev reflection, a gap is opened which is topologically different from the gap induced by direct tunneling between the edge states \cite{aliceaNew2012,beenakkerRandommatrix2015}. It has been predicted that a domain wall between these types of gapped edge states should host Majorana fermions in the case of the integer QH effect, and parafermions, a generalization of Majorana fermions, in the case of fractional QH states \cite{clarkeExotic2013,aliceaTopological2016}. These can be applied as topologically protected qubits \cite{kitaevFaulttolerant2003,lianTopological2018} and fractional transistors \cite{clarkeExotic2014}.

The experimental observations of signatures of Andreev reflection still require a deeper theoretical understanding. The experiments involve $s$-wave SCs, where Cooper pairing occurs between two electrons with opposite spins and zero total momentum. However, in the lowest QH edge state, spins are polarized due to the Zeeman effect, and still Andreev reflection is observed even for 2DEGs with no inherent spin-flip mechanisms (graphene in particular). In addition, the semiclassical picture involves an electron retroreflected as a hole, with no momentum transfer, while a more accurate quantum mechanical description of the problem treats the proximitized edge state as a hybridized particle-hole plane wave state \cite{hoppeAndreev2000,giazottoAndreev2005}. In this description, tunneling between the edge state and the bulk Cooper pair condensate should be heavily suppressed by conservation of momentum parallel to the interface; see \cref{fig:sketch}(b). In the absence of a strong spin-flip mechanism in the 2DEG, and for macroscopically long interfaces, observing Andreev reflection would then seem to break spin and momentum conservation.

In this paper we provide a resolution of these issues through a detailed analysis of a many-body tunneling model of a type-II SC tunnel coupled to a spin-polarized QH edge state. A spin-flip mechanism is provided at the SC surface in the form of Rashba spin-orbit coupling (SOC), and the screening current due to the Meissner effect in the SC is explicitly taken into account. By integrating out the SC, we find an effective 1D $p$-wave pairing Hamiltonian at low energies in the edge state. We show that the edge state can inherit SOC from the SC surface, which is of particular importance for the case of graphene, which has negligible SOC itself, as suggested in Refs.~\cite{leeInducing2017,gulInduced2020}. The proposed model also makes it clear that in the tunneling limit, the shift of Cooper pair momentum due to the screening current allows momentum-conserving chiral Andreev reflection in the edge state. The present model thus offers a qualitative explanation for the experimental observations of Refs.~\cite{zhaoInterference2020,hatefipourInduced2021}.

Our investigation of the tunneling limit differs from previous works \cite{hoppeAndreev2000,giazottoAndreev2005,vanostaaySpintriplet2011a}, where the system was modeled in the limit of a highly transparent interface, and Andreev reflection was possible due to evanescent single-electron excitations in the SC. With a many-body approach, we show that the low-transparency limit differs qualitatively from this case by the introduction of a resonance condition for tunneling. The systematic microscopic approach we follow provides the additional benefit that it allows for future extensions towards interacting electron systems such as the effective chiral Luttinger liquid edge state of fractional QH systems.

The structure of this paper is as follows: In Sec.~\ref{sec:model} we will describe the three individual ingredients of the system, namely, the SC, the QH system, and the tunneling Hamiltonian coupling them. In Sec.~\ref{sec:induced_SC}, we show how integrating out the SC gives rise to an induced pairing term and induced SOC in the QH system. In Sec.~\ref{sec:NSN} we analyze possible experimental signatures of the induced superconducting correlations. In Sec.~\ref{sec:discussion}, we discuss the main results, and in Sec.~\ref{sec:conclusions} we present our conclusions. Throughout the paper we use units of $\hbar = 1$.

\section{Model}\label{sec:model}

To investigate chiral Andreev reflection between a QH edge state and a SC, we consider an interface represented as a tunneling barrier along the $x$-axis between a semi-infinite SC at $y<0$ and a semi-infinite 2DEG in the $x-y$ plane at $y>0$; see~\cref{fig:sketch}. We apply a strong homogeneous magnetic field perpendicular to the QH plane with field strength $B_0$, which is below the critical field of the SC and falls off exponentially in the superconductor due to the Meissner effect. For simplicity we start by assuming that the SC is thick enough that no thin-film effects come into play, while discussion of more complicated cases follows later in the paper. The total magnetic field is then
\begin{align}
\v{B} = B_0 \Big[ e^{y/\lambda} \Theta(-y) + \Theta(y) \Big] \v{e}_z,
\label{eq:B}
\end{align}
where $\lambda$ is the London penetration depth, $\Theta$ is the Heaviside step function, and $\v{e}_z$ is the unit vector in the $z$ direction. The vector potential $\v{A}$ corresponding to the magnetic field must be continuous at the interface. It is convenient to choose the London gauge for the SC region, i.e., $\nabla \cdot \v{A} = 0$. In this gauge, $\v{A} \to 0$ inside the SC ($y \ll 0$), and $\v{A}$ has no normal component to the SC surface ($A_y=0$). Inside the 2DEG ($y > 0$) we choose the Landau gauge, $\v{A} = -B_0 (y+\lambda) \v{e}_x$ with $\v{e}_x$ parallel to the interface, which preserves translation invariance in $x$ and $z$, and where the addition of $\lambda$ ensures continuity at $y=0$. The total vector potential thus unifies the two gauges and remains continuous throughout space \cite{giazottoAndreev2005},
\begin{align}
\v{A}(y) = -B_0 \Big[ \lambda e^{y/\lambda} \Theta(-y) + (y + \lambda)\Theta(y)\Big] \v{e}_x.
\label{eq:A}
\end{align}
The London gauge \cite{londonProblem1948,bardeenChoice1951} allows for straightforward calculation of the screening current presented later in this paper.

\begin{figure}
\centering
\includegraphics{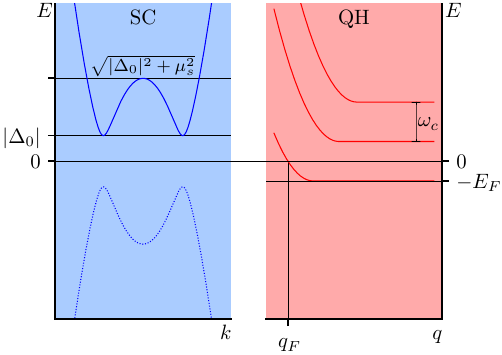}
\caption{A sketch of the relative energy scales of the SC and QH systems. All energies are measured with respect to the center of the SC gap. The relationships in the plot are not to scale, e.g. $|\Delta_0| \ll \mu_s$.}
\label{fig:Escales}
\end{figure}

We will now introduce the Hamiltonians for each part of the system, with all energies measured with respect to the center of the bulk SC gap. A reference sketch of the energy scales is provided in \cref{fig:Escales}.

\subsection{Quantum Hall edge state}

\begin{figure}[t]
\centering
\includegraphics{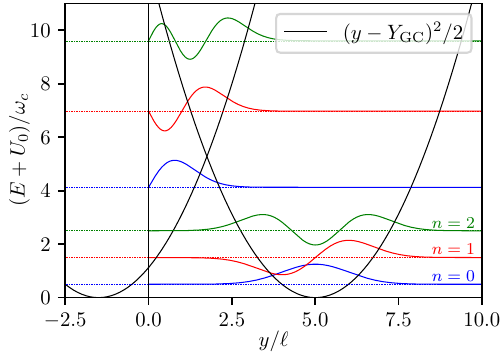}
\caption{The eigenstates of the QH Hamiltonian are harmonic oscillator states in the $y$ dimension in the bulk. The guiding center coordinate $Y_\text{GC}$ corresponds to the spatial minimum of the harmonic oscillator (solid black curves), and as it approaches and passes the hard wall at $y=0$ (solid black line), the states are squeezed and their energy is increased. Two examples are shown in the figure, an edge state with $Y_\text{GC}=-1.5 \ell$ to the left and a bulk state with $Y_\text{GC}=5\ell$ to the right. In each case the first three eigenstates are shown (solid colored curves) with dashed colored lines indicating the energy of the corresponding eigenstate. The eigenstates were found using the renormalized Numerov method.}
\label{fig:wall_states}
\end{figure}

The quantum Hall material in the $z=0, y>0$ region is described by a two dimensional electron gas (2DEG) in a homogeneous magnetic field, with Hamiltonian
\begin{align}
H_\text{QH} = \frac{1}{2 m_e}\Big(\v{q} + e \v{A}(y)\Big)^2 - U_0,
\label{eq:HQH}
\end{align}
where ${\v{q}} = (q_x, q_y)$ is the two-dimensional momentum operator of an electron in the 2DEG, $m_e$ is the electron mass and $e$ is the elementary charge. We have introduced the energy shift $U_0$ reflecting the band offset between the 2DEG and the SC. In the chosen Landau gauge, the eigenstates of this Hamiltonian are plane waves in the $x$ direction ($\propto e^{i q_x x}$ where $q_x$ is the canonical $x$-momentum), while in the $y$ direction they are harmonic oscillator eigenstates with eigenvalues $(1/2+n)\omega_c - U_0$ with non-negative integer $n$ and cyclotron frequency $\omega_c = eB_0/m_e$. These latter states are centered around the guiding center coordinate $Y_\text{GC} = q_x \ell^2 - \lambda$, corresponding to the harmonic oscillator minimum, with magnetic length $\ell = 1/\sqrt{eB_0}$. Due to translational invariance $q_x$ is a good quantum number, and the bulk dispersion is quantized into Landau levels, i.e., equidistant energy levels with macroscopic degeneracy in the quantum number $q_x$. At the edge of the QH material ($y \to 0^+$), the Landau levels bend upwards \cite{halperinQuantized1982} causing conducting edge states to propagate chirally around the edge of the material. This is caused by the squeezing of the bulk states induced by the termination of the system at $y=0$; see \cref{fig:wall_states}. The interface between a QH system and vacuum is conventionally modeled by hard-wall boundary conditions at $y=0$. We will adopt this assumption for the present calculation, and will discuss the consequences of more realistic, softer boundary conditions for the tunnel barrier to the SC in \cref{sec:discussion}.

\begin{figure}[t]
\centering
\includegraphics{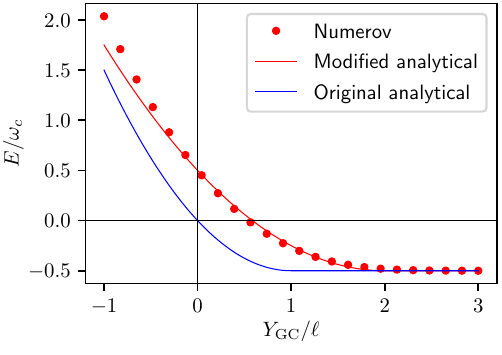}
\caption{The edge dispersion with $E_F=\omega_c/2$ of the lowest Landau level corresponding to the increase in energy due to eigenstates being squeezed against the wall; see \cref{fig:wall_states}. The red dots indicate the actual dispersion as calculated numerically using the renormalized Numerov method. The lower blue curve shows the approximate analytical expression of Ref. \cite{patlatiukEvolution2018}, which is accurate at higher energies, while the upper red curve shows the modified approximation in \cref{eq:1Ddisp}, which is more accurate for low energies.}
\label{fig:1DdispersionMod}
\end{figure}

A numerical solution for the dispersion of states near the hard wall was put forward in Ref.~\cite{meiHarmonic1983}, which we complement here by an adjusted version of the approximate analytic solution of Ref.~\cite{patlatiukEvolution2018} to be used for further analysis. We consider the lowest Landau level, $n=0$, and impose two conditions which determine the approximate edge dispersion up to a constant shift. These conditions lead to an analytical dispersion which reproduces numerical results to a good approximation. First, we assume that the dispersion is quadratic and that for $Y_\text{GC} \gg \ell$ we recover the bulk Landau levels. Second we require that in the bulk, $Y_\text{GC}> \ell$, the energy is $\omega_c/2- U_0$, while at the wall, $Y_\text{GC}=0$, the energy is $3 \omega_c/2 - U_0$, i.e. the ground state of the halved harmonic oscillator \cite{meiHarmonic1983}. These conditions yield the edge state dispersion of the lowest Landau level

\begin{align}
E_{q} &= \frac{\omega_c}{4 \ell^2}\Theta\Big(2 \ell - q \ell^2 + \lambda\Big) \Big(2 \ell - q\ell^2+\lambda \Big)^2 - E_F, \label{eq:1Ddisp}
\end{align}
where we used $q \equiv q_x$ as a shorthand, and have defined the Fermi energy $E_F$
\begin{align}
E_F &\equiv U_0 - \frac{\omega_c}{2},
\end{align}
which is tunable via the shift $U_0$. Fixing $U_0$ allows us to define the Fermi momentum $q_F$ in equilibrium by $E_{q_F}=0$, as well as the Fermi velocity $v_F$, both of which are functions of $E_F$,
\begin{align}
    q_F(E_F) &= \frac{2}{\ell} \left( 1 + \frac{\lambda}{2 \ell} -\sqrt{\frac{E_F}{\omega_c}} \right), \notag \\
    v_F(E_F) &= \partial_{q} E_q|_{q = q_F} = -\ell \sqrt{E_F \omega_c} = - \sqrt{\frac{E_F}{m_e}}.
\label{eq:EF}
\end{align}
Due to the choice of gauge in \cref{eq:A}, $\lambda$ appears in the QH edge spectrum. This reflects the gauge dependence of the canonical momentum, and ensures the gauge independence of the energy. Note that the gauge-independent quantities $E_F$ and $v_F$ are $\lambda$-independent, whereas the canonical Fermi momentum $q_F$ is gauge-dependent and thus can depend on $\lambda$. Compared with the approximative spectrum in Ref.~\cite{patlatiukEvolution2018}, we have replaced $\ell \to 2 \ell$ inside the square and the Heaviside function, as well as $\omega_c/2\ell^2 \to \omega_c/4\ell^2$. These substitutions provide a more accurate approximation for low energies. We compare the two approximations with numerical results found using the renormalized Numerov method in \cref{fig:1DdispersionMod}, and refer the reader to App.~\ref{app:numerov} for more information on the numerical method.

In the following, we will assume that the QH system is at filling factor $\nu = 1$, corresponding to a filled lowest Landau level and one edge channel. While in principle our model also applies to higher filling factors where interesting oscillatory phenomena occur \cite{hoppeAndreev2000,giazottoAndreev2005}, we note that the superconducting proximity effect can induce cross-channel interactions \cite{fisherCooperpair1994,michelsenCurrent2020} which are not captured by the presented model. We therefore limit the discussion to energies below the second Landau level. In the low-temperature limit and at $\nu=1$, we can model the edge channel as a non-interacting 1D system with Hamiltonian
\begin{align}
H_\text{edge} = \sum_q E_{q} \psi^\dagger_q \psi_q,
\label{eq:H1D}
\end{align}
where $E_{q}$ is defined in \cref{eq:1Ddisp} and $\psi_q$ is the annihilation operator for an edge electron with momentum $q$.

\subsection{Superconductor}
\label{sec:SC}
One of the main challenges for the experimental realization of the proposed system is to retain a superconducting state in a strong magnetic field. Experiments involve relatively thin films of dirty superconductors, such as NbN \cite{leeInducing2017} and MoRe \cite{ametSupercurrent2016,zhaoInterference2020}, whose superconducting parameters are intricate and very specific to the details of the material \cite{kamlapureMeasurement2010,sundarElectrical2013,makiseEstimations2018}. The fundamental physics of a conventional SC is however generic. In the interest of reaching an analytically tractable model, we have therefore decided to resort to a simplified model of the SC. While we should thus not expect this model to provide quantitative predictions, it will nonetheless give an accurate qualitative description of the system because the chiral Andreev reflection arises from the basic physical principles common to all superconductors.

We therefore start from a BCS Hamiltonian describing an $s$-wave SC in a magnetic field
\begin{multline}
H_\text{BCS} = \sum_\sigma \int d^3 r \Bigg[  c^\dagger_\sigma(\v{r}) \left( \frac{[-i \bm{\nabla} + e \v{A}(y)]^2}{2m_s} - \mu_s \right)  c_\sigma(\v{r})
\\\ - \left( \Delta_0  c^\dagger_\uparrow(\v{r})  c^\dagger_\downarrow(\v{r}) + \hc \right) \Bigg],
\label{eq:posBCS}
\end{multline}
where $ c_\sigma(\v{r})$ is the annihilation operator for an electron with spin $\sigma$ at position $\v{r} = (x,y,z)$, $m_s$ is the effective electron mass, $\mu_s$ is the chemical potential, and $\Delta_0$ is the superconducting order parameter in the form of a complex-valued constant. To highlight the role of the screening (Meissner) supercurrent induced by the magnetic field, we will formulate an equivalent momentum-space Hamiltonian where the magnetic field effect is entirely expressed in terms of the supercurrent. The current density can be expressed as $\v{j}_s = n_s e \v{v}_s$, where $n_s$ is the carrier density and $\v{v}_s = \v{k}_s/m_s$ is the carrier group velocity \cite{tinkhamUniform1996}. In the London gauge \cite{londonProblem1948} the current density is directly related to the vector potential through the London equation $\v{j}_s = - n_s e^2 \v{A}/m_s$, letting us identify $\v{k}_s = -e \v{A}$. Specifically, \cref{eq:A} implies
\begin{align}
\v{k}_s(y) = e B_0 \lambda e^{y/\lambda} \v{e}_x
\label{eq:momentum-assumption}
\end{align}
for $y \leq 0$. The current runs parallel to the interface along the $x$-axis with a magnitude which decays exponentially towards the SC bulk over a length given by the magnetic penetration depth $\lambda$. In the following, the amplitude of the supercurrent will matter for the tunnel coupling to the QH system. We see from \cref{eq:posBCS} that the effect of the magnetic field at the surface is a shift of the $x$ momentum. Gauging away this shift by transforming ${c}_\sigma(\v{r}) \to e^{i k_s(y) x} {c}_\sigma(\v{r})$, we see that such a shift is equivalent to multiplying a plane-wave phase factor by the order parameter $\Delta_0$ of the SC.

At energies within the SC gap, tunneling between the superconductor and the QH edge state is only possible by breaking a Cooper pair in the superconducting condensate. Since this costs an energy of order $\Delta_0$, the relevant length scale for tunneling is given by the superconducting coherence length $\xi$ \cite{larkinTheory2005}. Experiments on the proposed setup \cite{ametSupercurrent2016,leeInducing2017,zhaoInterference2020,hatefipourInduced2021} involve primarily type-II SCs where $\lambda \gg \xi$. For example, the ratio $\lambda/\xi$ is between 10 and 100 for thin films of NbN \cite{leeInducing2017,kamlapureMeasurement2010,makiseEstimations2018} and MoRe \cite{zhaoInterference2020,makiseEstimations2018,sundarElectrical2013}. Therefore, at the depths into which electrons can tunnel, we can neglect the $y$ dependence of the screening supercurrent, and thus approximate the vector potential by its constant value at the interface. This approximation allows us to simply use $e\v{A} = -k_s \v{e}_x$ with
\begin{align}
k_s = |\v{k}_s(y=0)| = e B_0 \lambda = \lambda/\ell^2.
\label{eq:ks}
\end{align}
Applying this approximation to \cref{eq:posBCS} and focusing on degrees of freedom of the superconductor within a distance $\xi$ from the QH system, we can describe the relevant SC surface states by the Hamiltonian,
\begin{multline}
H_{\text{BCS}}^{\text{surface}} = \sum_\sigma \int d^3 r \Bigg[  c^\dagger_\sigma(\v{r}) \left( -\frac{1}{2m_s} \bm{\nabla}^2 - \mu_s \right)  c_\sigma(\v{r})
\\\ - \left( \Delta_0 e^{- 2 i k_s x}  c^\dagger_\uparrow(\v{r})  c^\dagger_\downarrow(\v{r}) + \hc \right) \Bigg].
\label{eq:posBCS2}
\end{multline}
If we Fourier transform \cref{eq:posBCS2} and change variables to $\tvk = \v{k} - \v{k}_s$, we find
\begin{multline}
H_{\text{BCS}}^{\text{surface}} = \sum_{\tvk,\sigma} \left[\epsilon_{\v{k}_s+\tvk} {c}_{\v{k}_s+\tvk,\sigma}^\dagger {c}_{\v{k}_s+\tvk,\sigma}  \right.
\\
\left.+ \left( \Delta_0 {c}_{\v{k}_s+\tvk,\uparrow}^\dagger {c}_{\v{k}_s-\tvk,\downarrow}^\dagger + \hc \right) \right],
\end{multline}
where $\epsilon_\v{k} = \v{k}^2/2 m_s - \mu_s$. We conclude that under the assumption of negligible $y$-dependence of the screening current, the screening of the magnetic field in a type-II SC corresponds to pairing at the surface occurring between electrons with a total center of mass momentum of $\v{k}_s$.

For Andreev reflection to occur between an $s$-wave SC and a spin-polarized edge state, some spin-flip mechanism must be present. In experimental work on crossed Andreev reflection at low filling factors \cite{leeInducing2017,gulInduced2020}, the required spin-flip mechanism is proposed to be SOC in the SC material. While the geometry of these experiments does not directly correspond to that treated in this paper, we nevertheless expect SOC to be present at the surface of a high-field SC following the same arguments. We can then investigate the effect of this SOC in our model by considering Rashba SOC at the SC surface, which can be modeled by adding the term \cite{kimImpurityinduced2015}
\begin{align}
H_\text{SOC} = \alpha \sum_{\v{k}} (k_z + i k_x) {c}_{\v{k},\uparrow}^\dagger {c}_{\v{k},\downarrow} + \hc,
\end{align}
where $\alpha$ is the Rashba SOC strength. We then find the total SC surface Hamiltonian, valid within tunneling distance of the interface, to be given by
\begin{align}\label{eq:Hsurf}
H_\text{sc} &= H_{\text{BCS}}^{\text{surface}} + H_{\text{SOC}}  \\
&= \sum_{\tvk,\sigma} \left[\epsilon_{\v{k}_s+\tvk} {c}_{\v{k}_s+\tvk,\sigma}^\dagger {c}_{\v{k}_s+\tvk,\sigma} \right. \nonumber
\\
&\qquad \left. + \left( \Delta_0 {c}_{\v{k}_s+\tvk,\uparrow}^\dagger {c}_{\v{k}_s-\tvk,\downarrow}^\dagger + \hc \right) \right.
\nonumber
\\
&\qquad \left. + \left( \alpha [\tk_z + i (\tk_x+k_s)] {c}_{\v{k}_s+\tvk,\uparrow}^\dagger {c}_{\v{k}_s+\tvk,\downarrow} +\hc\right)  \right].\notag
\end{align}
For $\v{k}_s = k_s \v{e}_x$, we of course have $\tk_z = k_z$, but we maintain the above notation for the sake of consistency and generality. We note that earlier works have assumed the necessary spin-flip mechanism to be part of the tunneling Hamiltonian \cite{fisherCooperpair1994} or due to Rashba SOC in the 2DEG \cite{vanostaaySpintriplet2011a}. These approaches also allow Andreev reflection, but the present model has the benefits of applying equally to a 2DEG with negligible SOC such as graphene, as well as elucidating the role of SOC in the induced pairing after integrating out the SC, as will be shown in \cref{sec:induced_SC}.

We note that the derivation above assumes a simple quadratic dispersion of the SC material and requires a well-defined momentum. This can be a good approximation even for the complicated vortex structure of a type-II superconductor \cite{kohenProbing2006}. We note that the presence of vortices allows the magnetic field to penetrate the SC, decreasing the amplitude of the Meissner current at the surface and thus $k_s$. To describe superconductors with high critical fields, the parameters entering the Hamiltonian should thus not be regarded as microscopic parameters, but rather as effective parameters of the low-energy theory.

Previous works have shown that for a highly transparent interface, the screening current is not the only source of interface supercurrent \cite{hoppeAndreev2000,giazottoAndreev2005,vanostaaySpintriplet2011a}. If the coupling is strong, QH edge electrons penetrate into the SC in the form of an evanescent wave which is exponentially damped into the SC, but is a plane wave state parallel to the interface. This state has the same $x$ momentum as the edge state, and thus a supercurrent is induced by the proximity of the QH edge state. Our present model describes a qualitatively different situation, where the coupling between QH edge state and SC is weak, such that evanescence is negligible and the Meissner effect is the only source of surface supercurrent. The effect of evanescence could be taken into account by modifying $k_s$, but that is beyond the scope of this paper.

\subsection{Tunneling}
The origin of Andreev reflection is the tunneling of single electrons from the QH edge state into the SC near the interface. Weak tunneling across the interface can be described as single-electron tunneling between electronic states with identical spins states within a limited range in the $y$ and $z$ dimensions. Assuming the tunneling to be local in the fields $ \psi(x)$ and $ c_{\uparrow}(\v{r})$ the tunneling Hamiltonian is given by
\begin{align}
H_{\text{tunn}}
&= \gamma  \sqrt{w_z w_y} \int_{-L_{x}/2}^{L_{x}/2} dx\ [  \psi^\dagger(x)  c_{\uparrow}(x,0,0) + \hc] \nonumber
	\\
&= \Gamma \sum_{q,\v{k}} \delta_{q,k_x}  \left( {\psi}^\dagger_{q} {c}_{\v{k},\uparrow} + \hc \right),
\label{eq:HGamma}
\end{align}
where the second line is the Fourier transform of the position space expression. We have here defined
\begin{align}
\Gamma = \gamma \sqrt{\frac{w_z w_y}{L_z L_y}},
\label{eq:Gamma}
\end{align}
representing an effective tunneling amplitude with $\gamma$ being the associated energy, while $w_{y,z}$ are effective widths of the interface in the $y$ and $z$ directions, and $L_{y,z}$ are the lengths of the SC in the $y$ and $z$ directions. The Kronecker delta $\delta_{q,k_x}$ reflects the fact that local tunneling and translational invariance along the $x$ axis together imply conservation of $x$-momentum. Thus tunneling can only occur between electron states with the same momentum in the $x$ direction.

\section{Induced superconductivity}
\label{sec:induced_SC}

We now have a complete description of the system in the Hamiltonian
\begin{align}
H = H_\text{edge} + H_\text{sc} + H_{\text{tunn}},
\label{eq:fullH}
\end{align}
with the terms given in \cref{eq:H1D,eq:Hsurf,eq:HGamma}, respectively. Due to Andreev reflection an effective superconducting correlation will be induced into the QH edge state through the proximity effect. We will investigate this by integrating out the superconductor and finding an effective low-energy edge-state Hamiltonian for the hybrid Andreev edge state with superconducting pairing.

To diagonalize the SC surface Hamiltonian $H_\text{sc}$, we perform a Bogoliubov transformation and treat the SOC perturbatively to first order. Without SOC, the SC Hamiltonian is two-fold degenerate in spin, so we perform degenerate perturbation theory to first order in both energy and eigenstates. The resulting first order Hamiltonian
\begin{align}
H_\text{sc}^{(1)} = \sum_{\tvk} \sum_{j=\pm} \epsilon^{(1)}_{\tvk,j} {d}_{\tvk,j}^\dagger {d}_{\tvk,j}
\label{eq:Hsurf1}
\end{align}
has the eigenvalues
\begin{align}
\epsilon^{(1)}_{\tvk,\pm} &= \frac{\tk_x k_s}{m_s}  + \zeta_{\tvk} \notag \\
&\pm \alpha \left\vert k_s + (\tk_x+i \tk_z) \frac{\epsilon_{\tvk}+E_s}{\zeta_{\tvk} } \right\vert
\label{eq:SCspectrum}
\end{align}
where we have defined $E_s = k_s^2/2m_s$ and
\begin{align}
\zeta_{\tvk} &= \sqrt{\left(\epsilon_{\tvk}+E_s \right)^2 + |\Delta_0|^2}.
\end{align}
We have furthermore introduced the basis of the quasiparticle operators ${d}_{\tvk,j}$, which is defined explicitly below. It is important to note that the model depends on the assumptions that in comparison to the minimal excitation energy of the SC, the SOC amplitude is small enough to enable perturbative treatment, and the tunneling amplitude is small enough to neglect retardation when integrating out the SC; see \cref{app:action}. Since the minimal excitation energy depends on the first term of \cref{eq:SCspectrum}, we assume this shift to be not be too strong. In other words, we consider the case when the Meissner current is far from the critical current where the linear term of \cref{eq:SCspectrum}, the Doppler shift \cite{bagwellCritical1994}, closes the SC energy gap. We can then estimate the upper limit on $k_s$ as follows: Assuming $\alpha = 0$ and $|E_s| \ll \mu_s$, the new minimum excitation energy \cite{tinkhamUniform1996} is approximately given by
\begin{align}
\min\left(\epsilon^{(1)}_{\tvk,\pm}\big\vert_{\alpha=0}\right) = |\Delta_0| - \frac{|k_F| k_s}{m_s},
\end{align}
where $k_F^2/2m_s = \mu_s$. To remain below critical current we then must have
\begin{align}
k_s < \frac{|\Delta_0| m_s}{|k_F|}.
\end{align}
We will assume this condition to hold henceforth.

The quasiparticle operators ${d}_{\tvk,j}$ are related to the electron operators in the superconductor $c_{\v{k},\sigma}$ by
\begin{align}
& \begin{pmatrix}
{c}_{\v{k}_s+\tvk,\uparrow}\\
{c}_{\v{k}_s-\tvk,\downarrow}^\dagger
\end{pmatrix}
\label{eq:quasiparticleBasis} \\
& =
\frac{1}{\sqrt{2}}
\begin{pmatrix}
-u_{\tvk} & v_{\tvk}\\
v^*_{\tvk}& u^*_{\tvk}
\end{pmatrix}
\begin{pmatrix}
-i e^{-i\theta_{+}} {d}_{{\tvk},1}  + i e^{-i\theta_{+}}{d}_{{\tvk},2}\\
 e^{-i\theta_{-}}{d}_{-{\tvk},1}^\dagger +  e^{-i\theta_{-}} {d}_{-{\tvk},2}^\dagger
\end{pmatrix}, \notag
\end{align}
where
\begin{align}
\theta_\pm = \frac{1}{2}\text{Arg}\left(k_s \pm (\tk_x + i \tk_z) \frac{\epsilon_{\tvk}+E_s}{\zeta_{\tvk} } \right),
\end{align}
and $u_{\tvk}, v_{\tvk}$ are the Bogoliubov transformation parameters,
\begin{subequations}
\begin{align}
|u_{\tvk}|^2 = \frac{1}{2} \left(1+ \frac{\epsilon_{\tvk} + E_s}{\zeta_{\tvk}}\right),\\
|v_{\tvk}|^2 = \frac{1}{2} \left(1- \frac{\epsilon_{\tvk} + E_s}{\zeta_{\tvk}}\right),
\end{align}
\label{eq:uvNonEff}
\end{subequations}
with phases adding up to that of the superconducting order parameter, $\arg(u_{\tvk})+\arg(v_{\tvk})=\arg(\Delta_0)$.

Expressing the tunneling Hamiltonian in the quasiparticle basis leaves us with a total Hamiltonian which depends on the quasiparticle fields ${d}$ and the edge state electron fields ${\psi}$. To integrate out the superconductor, we translate this Hamiltonian to the corresponding Euclidian action
\begin{align}
S = S_\text{edge}[\psi,\psi^\dagger] + S^{(1)}_\text{sc}[ d, d^\dagger] + S_\text{tunn}[\psi,\psi^\dagger, d, d^\dagger].
\end{align}
Integrating out the superconductor corresponds to performing a functional integral over the Grassmann fields for the quasiparticles $d$, which yields an effective QH edge state action $S_\text{edge}^{\text{eff}} = S_\text{edge} + \delta S$ (see \cref{app:action} for details). This procedure shows that the tunneling gives rise to an effective retarded interaction within the QH edge state. In the low-energy limit, the retardation can be neglected and the effective action can be translated back to the corresponding Hamiltonian
\begin{multline}
H^\text{eff}_{\text{edge}} = \sum_{\tq} \Big[ (E_{k_s+ \tq} + \delta E_{\tq} )  \psi_{k_s+ \tq}^\dagger  \psi_{k_s+\tq} \\
+ \left(\Delta_{\tq} {\psi}^\dagger_{k_s+ \tq} {\psi}^\dagger_{k_s - \tq} + \hc \right)
\Big],
\label{eq:effH}
\end{multline}
where we have defined the momentum variable
\begin{align}
\tq = q - k_s,
\end{align}
which measures the edge electron momentum relative to $k_s$. In \cref{eq:effH}, $\delta E_{\tq}$ is a correction to the edge state dispersion and $\Delta_{\tq}$ is the effective pairing amplitude. We notice furthermore that the induced pairing in \cref{eq:effH} acts between electrons with total momentum $2k_s$ due to the presence of the Meissner current. Hence, we can expect that the effect of pairing is strongest for $q_F \approx k_s$. Using Eq.~(\ref{eq:EF}) and the definition of $k_s$ in \cref{eq:ks}, this entails
\begin{align}\label{eq:res}
    E_F \approx \omega_c,
\end{align}
which means that the Fermi level of the QH system should be close to the second Landau level; see Fig.~\ref{fig:Escales}.

In that case, the interplay between the edge states of the two Landau levels might become important, as has been shown using a phenomenological model \cite{fisherCooperpair1994,michelsenCurrent2020}. However, the resonance condition (\ref{eq:res}) is a consequence of the Landau level edge dispersion depicted in Fig.~\ref{fig:1DdispersionMod}. This spectrum has been derived under the assumption of hard-wall boundary conditions at $y=0$. In contrast, a smoother confining potential would generally lower the resonance energy to values $E_F < \omega_c$. To account for this effect, we will use the variable $\Omega$ to denote the offset between the center of the superconducting gap and the Fermi level of the QH edge state. For hard-wall boundary conditions, we have
\begin{align}
\Omega = \omega_c - E_F.
\end{align}

The total effective Hamiltonian is then a modified edge state Hamiltonian with added pairing between same-spin electrons at momenta $k_s \pm {\tq}$. We diagonalize it to find the spectrum of the edge states with induced superconductivity
\begin{align}
E^{\text{eff}}_{\tq,\pm} &=
\frac{1}{2} \left[ \tilde{E}_{{\tq}} - \tilde{E}_{-{\tq}} \pm \sqrt{ \left(\tilde{E}_{\tq}+\tilde{E}_{-{\tq}} \right)^2 +4 |\Delta_{{\tq}}|^2} \right],
\label{eq:effective_dispersion}
\end{align}
where
\begin{align}
\tilde{E}_{\tq} &= E_{k_s+{\tq}} + \delta E_{\tq},
\label{eq:tildeE}
\end{align}
and using Eq.~(\ref{eq:1Ddisp}),
\begin{align}
    E_{k_s + \tq} &= \frac{\omega_c \ell^2}{4}\Theta(2/\ell - \tq) \Big(2/\ell - \tq \Big)^2 - E_F \notag \\
&\approx
    \Omega + v_F \tq.
\end{align}
In the last line we expanded the energy to first order in $\delta q$ and introduced the Fermi velocity $v_F \equiv v_F(\omega_c) = -\omega_c \ell = -\sqrt{e B_0}/m_e$ [see Eq.~(\ref{eq:EF})]. In the next section we will present the exact expressions for $\Delta_\tq$ and $\delta E_\tq$.

\subsection{Pairing and dispersion}
\label{sec:pairing-and-dispersion}

\begin{figure}
\includegraphics{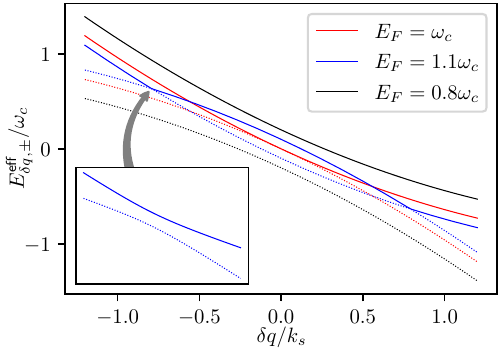}
\caption{The effective dispersion in \cref{eq:effective_dispersion} for different Fermi energies, with linearized versions of $\delta E_\tq$ and $\Delta_\tq$. Solid and dotted curves denote particle-like ($E^{\text{eff}}_{\tq,+}$) and hole-like ($E^{\text{eff}}_{\tq,-}$) bands, respectively. The inset shows the avoided crossing for $E_F>\omega_c$. The dispersion has been calculated for $m_s = 10 m_e$, $\mu_s = 0.4 \times 10^4 \omega_c$, $|\Delta_0| = 40 \omega_c$, $\lambda = 0.8 \ell$, and $\Gamma_L = \alpha = \omega_c \ell$.}
\label{fig:EFrow}
\end{figure}

Diagonalizing the effective edge state Hamiltonian lets us express the induced pairing contribution to this Hamiltonian analytically. Due to the fermionic nature of the fields $\Delta_{{\tq}}$ is an odd function in ${\tq}$ and we can express it as $\Delta_{{\tq}} = (\tilde{\Delta}_{{\tq}}-\tilde{\Delta}_{-{\tq}})/2$, with
\begin{align}
\tilde{\Delta}_{{\tq}} = i |\Delta_0| \Gamma^2 \alpha \sum_{k_y,k_z}
\frac{k_s \zeta_{\vkappa} + (\epsilon_\vkappa + E_s) {\tq}}{\zeta_{\vkappa}^2 ({\tq} k_s/m_s + \zeta_{\vkappa})^2},
\label{eq:deltaI}
\end{align}
where $\vkappa = ({\tq},k_y,k_z)$ and we have ignored terms of order $\mathcal{O}(\alpha^2)$ to be consistent with the perturbation calculation above. Note that since this effective pairing is the result of integrating out the SC, it only depends on parameters of the SC and the tunneling Hamiltonian, as well as the dimensionality of the edge system. The same result can thus be extended to induced pairing in any 1D system.

For small momenta ${\tq}$ we can describe the induced pairing as only the linear ($p$-wave) component
\begin{align}
\Delta_{{\tq}} \simeq \frac{i}{2} \DeltaP \tq,
\label{eq:linDelta}
\end{align}
with $\DeltaP = -2 i \partial_{\tq} \Delta_{{\tq}} \vert_{{\tq}=0} \in \mathbb{R}$. Assuming macroscopic dimensions $L_y, L_z$ of the SC allows us to approximate the sum in \cref{eq:deltaI} with an integral, which leads to
\begin{multline}
\DeltaP = \frac{2 \Delta_0 m_s \alpha \Gamma_L^2}{\pi |\Delta_0|^3}
 \Bigg\{ |\Delta_0| \left( \frac{ |\Delta_0|^2/4 + E_s (E_s-\mu_s )}
{|\Delta_0|^2 + ( E_s - \mu_s )^2} \right)
\\
+ E_s \Big[ \arctan\Big( \frac{E_s - \mu_s}{|\Delta_0|}\Big) - \frac{\pi}{2} \Big] \Bigg\},
\label{eq:effDelta}
\end{multline}
where $\Gamma_L = \Gamma \sqrt{L_z L_y}$. This can be simplified further by noting that in a metallic superconductor we have $E_s, |\Delta_0| \ll \mu_s$, leading to the approximate expression
\begin{align}
\DeltaP \simeq - 2 \Delta_0 m_s \alpha \Gamma_L^2 \frac{E_s}{|\Delta_0|^3}.
\end{align}
This expression makes it clear that a nonzero induced pairing amplitude requires Rashba SOC ($\alpha \neq 0$), a Meissner supercurrent ($E_s \neq 0$), and electron tunneling ($\Gamma_L\neq 0$). Note that the induced pairing amplitude $\Delta_{{\tq}}$ in the low energy theory involves a factor of $1/|\Delta_0|$ which is proportional to the coherence length $\xi$ of a Cooper pair in the SC. This is a consequence of the induced $p$-wave pairing in a chiral edge state $\psi(x)\psi(x')\simeq (x'-x)\psi(x)\partial_x\psi(x)\simeq\xi\psi(x)\partial_x\psi(x)\propto\Delta_0^{-1} \psi(x)\partial_x\psi(x)$.

In addition to the induced pairing, the proximity effect modifies the kinetic energy by the shift
\begin{align}
\delta E_{\tq} = 2 \Gamma^2 \sum_{k_y,k_z} \frac{\epsilon_\vkappa + E_s - \tq k_s/m_s}{\zeta^2_\vkappa - ({\tq} k_s/m_s)^2}.
\label{eq:dE}
\end{align}
Taking the small-${\tq}$ limit as above, we find $\delta E_{\tq} = \delta E_0 + \delta v \tq$, where
\begin{align}
\delta E_0 &= \delta E_{\tq=0},
\end{align}
and
\begin{align}
\delta v &= \partial_{\tq} \delta E_{\tq} \vert_{{\tq}=0}
\end{align}
Explicitly, we find
\begin{align}
\delta v &= -\frac{k_s \Gamma_L^2}{\pi |\Delta_0|} \left[ \frac{\pi}{2} - \arctan\left( \frac{E_s-\mu_s}{|\Delta_0|} \right) \right] \simeq -\frac{k_s \Gamma_L^2}{|\Delta_0|}
\label{eq:effdE}
\end{align}
where we used the approximation $E_s, |\Delta_0| \ll \mu_s$ for the final expression. This should be regarded as a renormalization of the Fermi velocity $v_F$ defined in Eq.~(\ref{eq:EF}) due to the tunneling to the superconductor.

The induced constant shift $\delta E_{0}$ formally diverges logarithmically due to the infinite limits on the sum in \cref{eq:dE}; however, in practice we only sum up to the SC bandwidth, and we can thus absorb $\delta E_{0}$ into the $U_0$ shift and ignore it throughout the rest of the paper. We note that the kinetic energy modification $\delta E_{\tq}$ depends on the supercurrent but does not require SOC. This reflects the fact that the supercurrent allows tunneling to occur, while the SOC allows the spin singlet pairing of the SC to induce spin triplet pairing in the QH edge.

Using \cref{eq:effDelta,eq:effdE} we plot the effective dispersion given by \cref{eq:effective_dispersion} in \cref{fig:EFrow}. The dispersion consists of two bands which are similar to the particle-hole bands of a QH edge without induced superconductivity, but the corresponding states are now hybrid particle-hole states. We note that for $E_F>\omega_c$ the bands have an avoided crossing, while for $E_F< \omega_c$ they do not.

\begin{figure}
\centering
\includegraphics[width=0.9\columnwidth]{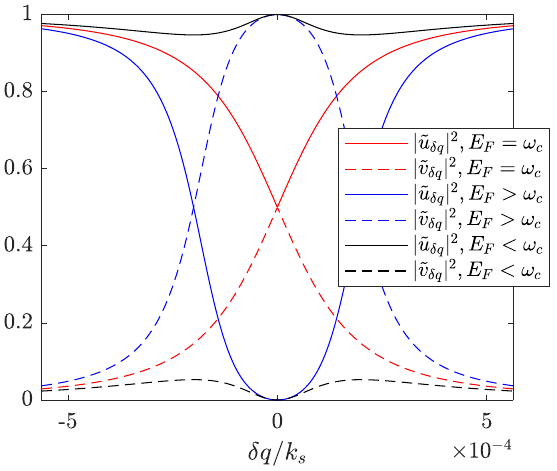}
\caption{The amplitudes of the particle state $|\tilde u_\tq|^2$ (solid curves) and the hole state $|\tilde v_\tq|^2$ (dashed curves) in the eigenstates of the effective Hamiltonian, as given by \cref{eq:uvEff}. For illustrative purposes, the deviations from the identity are chosen to be small, i.e., $|E_F/\omega_c - 1|\sim 10^{-9}$. The parameters are the same as in \cref{fig:EFrow}.}
\label{fig:uvplot}
\end{figure}

The eigenstates of the effective Hamiltonian in \cref{eq:effH} are linear combinations of particle and hole states, with amplitudes given by the Bogoliubov transformation parameters
\begin{subequations}
\begin{align}
|\tilde{u}_{\tq}|^2 = \frac{1}{2} \left(1+ \frac{\tilde{E}_{\tq}+\tilde{E}_{-\tq}}{\left(\tilde{E}_{\tq} + \tilde{E}_{-\tq}\right)^2 + 4 |\Delta_\tq|^2}\right),\\
|\tilde{v}_{\tq}|^2 = \frac{1}{2} \left(1- \frac{\tilde{E}_{\tq}+\tilde{E}_{-\tq}}{\left(\tilde{E}_{\tq} + \tilde{E}_{-\tq}\right)^2 + 4 |\Delta_\tq|^2}\right),
\end{align}
\label{eq:uvEff}
\end{subequations}
where $|\tilde{u}_{\tq}|^2$ is the amplitude of the particle state and $|\tilde{v}_{\tq}|^2$ is the amplitude of the hole state. The energy $\tilde E_{\tq}$ is defined in \cref{eq:tildeE}. These are shown in \cref{fig:uvplot} for various values of $\Omega = \omega_c - E_F$, and we see that at $\Omega = 0$ the system undergoes a phase transition, as originally described in Ref.~\cite{readPaired2000}. In the next section, we will investigate the impact of this transition point on transport through the system.

\section{Normal-superconductor-normal junction}
\label{sec:NSN}

\begin{figure}
\includegraphics[width=\columnwidth]{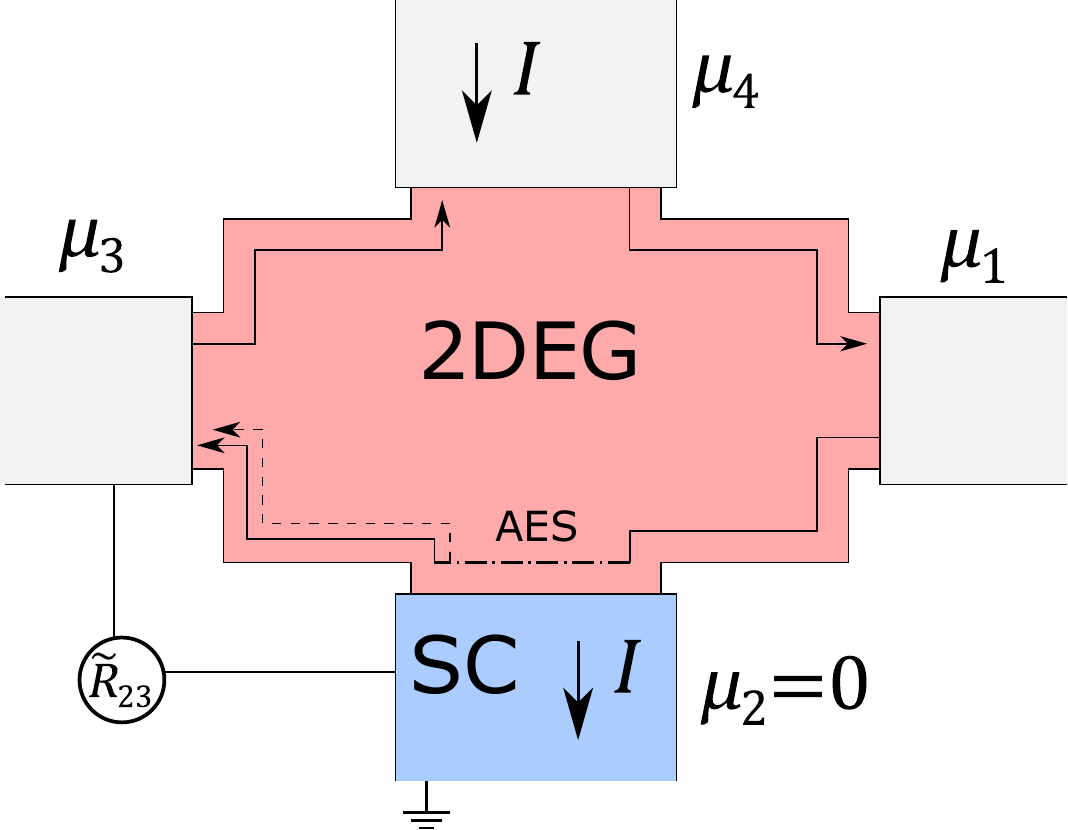}
\caption{A sketch of an experimental measurement of the downstream resistance $\tilde R_{23} = dV_3/dI$, where $V_3 = \mu_3/e$. A current $I$ is injected from lead 4 to 2, while the two voltage probes (left and right) measure $\mu_1$ and $\mu_3$. Measuring $\tilde R_{23} < 0$ indicates the presence of the Andreev edge state (AES). Note that the choice of energy gauge in this paper leads to $\mu_2=0$.}
\label{fig:RxxSketch}
\end{figure}

A useful experimental signature of induced superconductivity in a chiral edge state is the resistance measured across a superconducting electrode interfaced with the 2DEG. The induced superconducting correlations correspond to a mixing of electron and hole states, leading to a finite probability of an electron being converted into a hole as it passes the SC. This can be detected experimentally as a negative effective resistance \cite{zhaoInterference2020,hatefipourInduced2021} for the geometry treated below, as well as for geometries allowing crossed Andreev reflection \cite{leeInducing2017,gulInduced2020}.

Let us consider the four-terminal system of \cref{fig:RxxSketch} at filling $\nu=1$, and with $\mu_2=0$. We will first derive the transfer matrices across the normal-superconductor-normal (NSN) junction, which lets us define the probability $P_h$ of an electron being transmitted across the junction as a hole in a process similar to crossed Andreev reflection. Due to the chirality of the edge state, there is no backscattering and thus no Andreev reflection within the edge state. We will then investigate the four-terminal scattering problem to relate $P_h$ to the differential downstream resistance $\tilde R_{23} = dV_3/dI$, where $V_3 = \mu_3/e$.

The starting point of the scattering calculation is the Hamiltonian in \cref{eq:effH}, which governs the QH edge state with induced superconductivity. To simplify the scattering calculation, we linearize the Hamiltonian in $|\delta q| \ll k_s$, such that
\begin{align}\label{eq:Heff_q}
    H_\text{edge}^{\text{lin}}
&=
    \frac{1}{2} \sum_{\tq} \begin{pmatrix} p^\dag_\tq & h^\dag_\tq \end{pmatrix} \left[ M(\tq) + \Omega \sigma_z \right]
    \begin{pmatrix} p_\tq \\ h_\tq \end{pmatrix},
\end{align}
with
\begin{align}
M(\delta q)
&=
    \begin{pmatrix}
        v_F + \delta v & i\DeltaP \\
        - i\DeltaP & v_F + \delta v
    \end{pmatrix} \tq,
\end{align}
where we assumed $\DeltaP$ to be positive without loss of generality and where $\sigma_z$ is a Pauli matrix. Moreover, we introduced Nambu spinors consisting of the modes $p^\dag_{\tq} =  \psi^\dag_{k_s+\tq}$ and $h^\dag_{\tq} =  \psi_{k_s-\tq}$. We note that this further linearization does not account for the avoided crossings seen in \cref{fig:uvplot}, and so moving forward we consider only the case $\Omega > 0$, i.e., $\omega_c > E_F > 0$. This corresponds to the case where the quantum Hall system is in the lowest Landau level.

In order to determine the scattering matrix for transport from lead 1 to lead 3 of \cref{fig:RxxSketch}, we consider an infinite 1D edge state along the $x$ axis subject to induced superconductivity over a finite length $L$. For this purpose, we need to construct a position-space representation of Eq.~(\ref{eq:Heff_q}). Fourier transforming (\ref{eq:Heff_q}) and taking into account Hermiticity, one finds the effective Hamiltonian for the NSN junction,
\begin{align}
  H_{\NSN} &=
    - \frac{i}{4} \int dx \begin{pmatrix} p^\dag(x) & h^\dag(x) \end{pmatrix} \{ M(x),  \partial_x\}
    \begin{pmatrix} p(x) \\ h(x) \end{pmatrix} \notag \\
  &+ \frac{\Omega}{2} \int dx \begin{pmatrix} p^\dag(x) & h^\dag(x) \end{pmatrix} \sigma_z\begin{pmatrix} p(x) \\ h(x) \end{pmatrix} , \label{eq:HNSN}
\end{align}
where $\{\cdot,\cdot\}$ is the anticommutator and
\begin{align}
  M(x) &= \begin{pmatrix}
      v_F(x) & i \DeltaP(x) \\
      -i \DeltaP(x) & v_F(x)
\end{pmatrix}.
\end{align}
In the superconducting region ($x \in [-L/2,L/2]$), we assume that $\DeltaP(x) \equiv \DeltaP = \text{const.}$, while $\DeltaP(x) = 0$ otherwise. Moreover, we assume that $|\DeltaP(x)| < |v_F(x)|$ for all $x$, which makes $M(x)$ positive definite. In this limit we can, for clarity, consider the Fermi velocities in the normal and superconducting sections as equal, i.e., $v_F(x) = v_F = \text{const}$. This can be done without loss of generality since the results turn out to only depend on the value of $v_F(x)$ in the superconducting region due to the lack of backscattering modes.

The problem has now been cast in the form of a Bogoliubov-{de Gennes} Hamiltonian, with which we can find the energy eigenstates for a spatially inhomogeneous superconducting system \cite{vanostaaySpintriplet2011a}. By integrating the single-particle Schr\"odinger equation $\mathcal{H} \phi(x) = E\phi(x)$ corresponding to \cref{eq:HNSN} over small intervals near the interfaces at $x = \pm L/2$, one can derive the boundary conditions $A(x^+) \phi(x^+) = A(x^-) \phi(x^-)$ where $A(x) = \sqrt{M(x)}$ and the superscripts $+$ and $-$ denote the position infinitesimally far to the right and left of the discontinuity, respectively. As $A(x)$ is discontinuous near these points, $\phi(x)$ is discontinuous at the interfaces as well. We solve the single-particle Schr\"odinger equation in each region and use the boundary condition to connect the solutions. This allows us to derive the transfer matrix $T_\text{SN}$ relating the scattering states on both sides of a given interface, e.g., $\phi(L/2^-) = T_\text{SN} \phi(L/2^+)$, as well as the transfer matrix across the superconducting region, $\phi(-L/2^+) = T_\text{SS} \phi(L/2^-)$. Taking into account the boundary conditions, we find the total transfer matrix for the full system $\phi(-L/2^-) = T_\text{NSN} \phi(L/2^+)$ for a scattering state with energy $E$ to be
\begin{align}
&T_\text{NSN} =
\begin{pmatrix}
t_{pp} & t_{ph}\\
t_{hp} & t_{hh}
\end{pmatrix} =
    e^{-2 i v_F E L/\tilde{v}_F^2}  \\
&\times \begin{pmatrix}  \cos\left(\frac{L}{\tilde{L}}\right) + \frac{i \Omega \tilde{L}}{\tilde{v}_F} \sin\left(\frac{L}{\tilde{L}}\right) &  - \frac{2 \DeltaP E \tilde{L}}{\tilde{v}_F^2} \sin\left(\frac{L}{\tilde{L}}\right)  \\ \frac{2 \DeltaP E \tilde{L}}{\tilde{v}_F^2} \sin\left(\frac{L}{\tilde{L}}\right) & \cos\left(\frac{L}{\tilde{L}}\right) - \frac{i \Omega \tilde{L}}{\tilde{v}_F} \sin\left(\frac{L}{\tilde{L}}\right)  \end{pmatrix},\notag
\end{align}
where $t_{hp}$ is the amplitude of a particle being transmitted as a hole, and we have defined the parameters
\begin{align}
\tilde{v}_F &= \sqrt{v_F^2-\DeltaP^2},\\
\tilde{L} &= \frac{\tilde{v}_F^2}{\sqrt{ 4 \DeltaP^2 E^2 + \tilde{v}_F^2 \Omega^2}}.
\end{align}
From the scattering matrix, we can extract the probability of converting an electron entering the NSN junction at position $x=L/2^+$ into a hole exiting it at position $x=-L/2^-$ as $P_h = |t_{hp}|^2$,
\begin{align}
P_h(E) &= \frac{1}{1 + \frac{1}{4} (\Omega/E)^2[(v_F/\DeltaP)^2-1]} \label{eq:Ph} \\
&\times \sin^2\left( \frac{\sqrt{ 4 + (\Omega/E)^2[(v_F/\DeltaP)^2-1]}}{(v_F/\DeltaP)^2-1} \frac{E L}{\DeltaP} \right).
\notag
\end{align}
Due to unitarity, we have $P_p = |t_{pp}|^2 = 1 - P_h$. The expression consists of a Lorentzian prefactor and a squared sine factor which depends on the length of the interface $L$. The oscillating part describes Fabry-P\'erot resonances due to repeated Andreev reflections inside the superconducting region. As discussed in Refs.~\cite{zhaoInterference2020,hatefipourInduced2021}, the interference causing the oscillation can be understood by expressing the hole transmission amplitude as $P_h(E) \propto \sin^2[q_\text{ph}(E) L/2]$, where $q_\text{ph}(E)$ is the momentum difference between the particle-like and hole-like states, both with energy $E$, propagating through the SC region.

The Lorentzian prefactor is independent of $L$ and describes the maximum electron-hole conversion rate. It vanishes $\propto (\DeltaP/v_F)^2$ for small $\DeltaP$. Moreover, assuming that $\DeltaP \ll v_F$ it vanishes $\propto (\DeltaP E/v_F\Omega)^2$ at small energies $|E| \ll |\Omega|$. This is a consequence of Pauli blocking \cite{fisherCooperpair1994,virtanenSignatures2012,michelsenCurrent2020} as the $p$-wave pairing $\DeltaP$ cannot act between two electrons at $E=0$. On the other hand, the prefactor approaches unity $\propto 1 - (v_F \Omega/\DeltaP E)^2$ at energies $|E| \gg |\Omega|$. In this case, an incoming electron with energy $E$ can in principle be paired with an electron with energy $-E$. However, this reasoning breaks down when $|E|$ approaches $|\Delta_0|$. Whether the pairing takes place also depends on the availability of electrons at these energies, which will be accounted for by the Fermi functions which will appear in the following calculation of the current.

An experimentally accessible quantity which carries information about $P_h(E)$ is the differential downstream resistance $\tilde R_{23}$ in the four-terminal setup shown in \cref{fig:RxxSketch}. Here, terminals $1$ and $3$ act as voltage probes (i.e., $\braket{I_{1,3}} = 0$) with respective chemical potentials $\mu_{1,3}$. Moreover, a current $I$ is injected via terminal $4$ and, due to current conservation, leaves the setup via the grounded superconductor at terminal $2$. One can extend the Landauer-B\"uttiker formalism \cite{dattaElectronic1995} to account for both particles and holes (see App.~\ref{app:LBextended}). In the basis of incoming and outgoing particles and holes at leads $1$, $3$ and $4$, the unitary scattering matrix reads
\begin{align}
    T = \begin{pmatrix}
        0 & 0 & \mathbbm{1}_2 \\
        T_{\rm NSN} & 0 & 0 \\
        0 & \mathbbm{1}_2 & 0
    \end{pmatrix}.
\end{align}
Using the Landauer-B\"uttiker formula (\ref{eq:LB_Im}) and assuming zero temperature to replace the Fermi functions by step functions, we arrive at the expression for the current in the different terminals,
\begin{align}
\braket{I_1} &= \frac{1}{2\pi} \left( \mu_1 - \mu_4 \right), \notag \\
\braket{I_3} &= \frac{1}{2\pi} \int_0^\infty dE \Big[ \Theta(-E+\mu_3)- \Theta(-E-\mu_3) \notag \\&+ [1-2P_h(E)] \Big(\Theta(-E-\mu_1)  - \Theta(-E+\mu_1) \Big)\Big], \notag \\
\braket{I_4} &= \frac{1}{2\pi} \left( \mu_4 - \mu_3 \right).
\label{eq:IwithE}
\end{align}
Since terminals 1 and 3 are voltage probes we have $\braket{I_{1,3}}=0$, while the injected current leads to $\braket{I_4} = I$. These constraints allows us to express all chemical potentials as a function of the injected current $I$
\begin{align}
I &= \frac{1}{\pi} \int_0^{\mu_1(I)} dE P_h(E), \label{eq:muI} \\
\mu_3 &= \mu_1(I) - 2\pi I. \notag
\end{align}
A few re-arrangements then allow us to finally evaluate the differential resistance
\begin{align}
\tilde R_{23} = \frac{1-2 P_h[\mu_1(I)]}{2 P_h[\mu_1(I)]}\frac{h}{e^2},
\label{eq:dVdI}
\end{align}
where we have restored physical units with $h$ being the Planck constant and $e$ being the elementary charge. This is the energy-dependent extension of the corresponding expression derived in Ref.~\cite{zhaoInterference2020}.

In general, it is not possible to determine $\mu_1(I)$ analytically from Eq.~(\ref{eq:muI}). However, to linear order in the injected current $I$, we can expand the integral and Eq.~(\ref{eq:dVdI}) to find
\begin{align}
\tilde R_{23} \propto \left[ P_h''(0) I^2 \right]^{-1/3}.
\end{align}
The peculiar scaling of the differential resistance as a function of bias current is due to the $p$-wave pairing which enforces $P_h(0) = 0$ as well as to the particle-hole symmetry which leads to $P_h(E) \propto E^2$ for small energies.

\begin{figure}[t]
\centering
\includegraphics[width=\columnwidth]{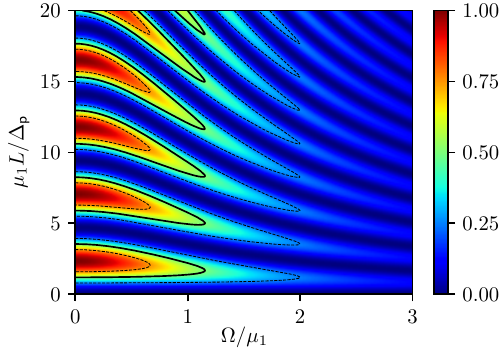}
\caption{The hole transmission probability $P_h$ with $|v_F/\DeltaP| = 2$ as a function of the system length $L$ (in units of $\DeltaP/\mu_1$) and the detuning from resonance $\Omega$ (in units of $\mu_1$). Negative resistance occurs for $P_h>1/2$ and the $P_h=1/2$ contours are marked by solid black curves. Note that we assume $\Omega >0$ in this plot as this corresponds to the lowest Landau level.}
\label{fig:Ph-mu1-plot}
\end{figure}

Experimentally a negative downstream resistance $\tilde R_{23} < 0$ is used as the primary indicator of induced superconducting correlations \cite{zhaoInterference2020,hatefipourInduced2021}. We see from \cref{eq:dVdI} that a negative resistance occurs when $P_h > 1/2$. In \cref{fig:Ph-mu1-plot} the hole transmission probability $P_h(\mu_1)$ is plotted as a function of $\Omega/\mu_1$ and $\mu_1 L/\DeltaP$. The contours where $P_h=1/2$ are marked by thick black curves. We immediately see that there are several islands of negative resistance. The behavior of $P_h$ along the line $\Omega/\mu_1$ is determined by the Lorentzian prefactor in \cref{eq:Ph}. In particular, no negative resistance is found beyond the half width at half maximum (HWHM) of this Lorentzian, which occurs at
\begin{align}
\left\vert\frac{\Omega_{\text{HWHM}}}{\mu_1}\right\vert = \frac{2}{\sqrt{(v_F/\DeltaP)^2-1}}.
\end{align}
Along the $\mu_1 L/\DeltaP$ axis, hole transmission peaks occur at the values
\begin{align}
\frac{\mu_1 L}{\DeltaP} =
\frac{\pi(2 n +1)[(v_F/\DeltaP)^2-1]}{2\sqrt{4 + (\Omega/\mu_1)^2[(v_F/\DeltaP)^2-1]}},
\end{align}
with integer $n$. At resonance, $\Omega=0$, the induced correlations are strong and perfect hole conversion $P_h=1$ becomes possible despite the model being in the tunneling limit, corresponding to low interface transparency.

\section{Discussion}
\label{sec:discussion}

\begin{figure}
\centering
\includegraphics[width=0.9\columnwidth]{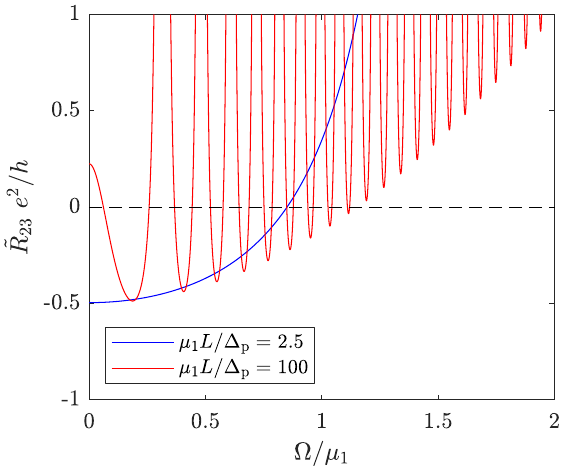}
\caption{For a fixed $|v_F/\DeltaP| = 2$, we predict increased resistance oscillations with increased $\mu_1 L/\DeltaP$, here shown with two examples of $\tilde R_{23}(\Omega/\mu_1)$ at fixed values of $\mu_1 L/\DeltaP$. At $\mu_1 L / \DeltaP = 2.5$ we have slow oscillations, i.e. not a full oscillation within the negative resistance region, while for $\mu_1 L/\DeltaP = 100$ we have fast oscillations.}
\label{fig:flatVsOsc}
\end{figure}

The work presented in this paper constitutes an analytical prediction of the behavior of induced superconductivity in the QH edge. It furthermore complements related work in taking into account the SOC originating in the SC, rather than in the 2DEG \cite{vanostaaySpintriplet2011a} or as an abstract spin-flip mechanism at the interface \cite{fisherCooperpair1994,michelsenCurrent2020}. It is expected that considering the SOC as originating in the SC corresponds well with experiments using Nb-based SC's \cite{wakamuraSpin2014}; see e.g. Ref.~\cite{leeInducing2017,gulInduced2020}.

In \cref{sec:NSN} we predicted negligible hole conversion in the limit $\mu_1=0$ due to Pauli blockade as well as due to the $p$-wave nature of the induced pairing, and we only find $P_h>1/2$ and thus negative resistance away from this limit. We note that previous experimental results \cite{zhaoInterference2020} include significant hole conversion in the $\mu_1=0$ limit which can be explained by the experiment being at higher filling factors which avoids the Pauli blockade.

We also predict that Andreev reflection at the interface is strongly suppressed away from the resonance condition $\Omega = 0$. While our model only explicitly concerns filling factor $\nu=1$, it can be most straightforwardly extended to higher filling factors by adding higher Landau levels while ignoring cross-interactions between the channels. The resonance condition would then occur once per Landau level. Previous numerical work \cite{giazottoAndreev2005} has found that Andreev reflection can occur at any energy across the Hall plateaus for high transparency, while lowering the transparency suppresses the process except at one energy per Hall plateau. While further investigation of this result would be needed, the resonance condition predicted in this paper is in agreement with the tendency indicated in the numerical results. Our results thus also suggest that varying the transparency of the interface may account for the difference between observing patches of induced superconductivity on a range of filling factors such as in Ref.~\cite{ametSupercurrent2016}, and observing induced superconductivity at a continuous range of filling factors such as in Refs.~\cite{zhaoInterference2020,hatefipourInduced2021}.

It is worth noting that in the presented model the resonance condition allowing strong Andreev reflection occurs when the edge state Fermi energy approaches the second Landau level from below. This is a problematic limit due to the appearance of the next Landau level and the ensuing cross-channel interactions. However, this resonance condition was derived for hard-wall boundary conditions, whereas more realistic soft boundaries lower the resonance energy and bring it between Landau levels. This is consistent with Ref.~\cite{giazottoAndreev2005}, which indicates that for higher interface transparency the resonance energy shifts downwards.

As illustrated in \cref{fig:flatVsOsc}, for a fixed ratio $v_F/\DeltaP$, the downstream resistance exhibits fast (slow) oscillations for high (low) values of $\mu_1 L/\DeltaP$. As discussed in \cref{sec:NSN}, this reflects the oscillations of $P_h(\mu_1) \propto \sin^2[q_\text{ph}(\mu_1) L/2]$ with the particle-hole state momentum difference
\begin{align}
q_\text{ph} = 2 \frac{\sqrt{ 4 + (\Omega/\mu_1)^2[(v_F/\DeltaP)^2-1]}}{(v_F/\DeltaP)^2-1} \frac{\mu_1}{\DeltaP}.
\end{align}
We note that this prediction represents a rigorous derivation of the models applied in the analysis of the experiments presented in Refs.~\cite{zhaoInterference2020,hatefipourInduced2021}. These experiments work with comparable geometries but different materials. One experiment observes strong oscillations for an interface length $L = 600\ \si{nm}$ \cite{zhaoInterference2020}, while the other one observes no oscillations for an interface length $L = 150\ \si{\mu m}$ \cite{hatefipourInduced2021}. Within the context of the presented theory, this would correspond to a large ratio $q_\text{ph,slow}/q_\text{ph,fast} \gg 1$ between the slow and fast oscillation experiments, respectively. With the results presented in \cref{sec:induced_SC}, this ratio can be directly related to experimentally applied parameters and material qualities.

As previously noted, the induced pairing results are in principle independent of the details of the edge state Hamiltonian as long as it is one-dimensional. Since the model is of a many-body nature, we are free to replace the edge state Hamiltonian with e.g.\ a chiral Luttinger liquid Hamiltonian, which describes a system of strongly interacting electrons in 1D. This would effectively allow us to model the case of induced pairing in a fractional QH edge state, a system which has recently become experimentally available \cite{gulInduced2020}.

We have shown that the existence of a surface supercurrent is essential for electrons to tunnel in a momentum-conserving system. The results of \cref{sec:induced_SC} do not depend on our assumption that the supercurrent is entirely due to the Meissner effect. For instance, to higher orders in tunneling, the tunneling electrons themselves could cause a nonzero supercurrent in an evanescent mode into the superconductor. One can account for such effects by regarding $k_s$ as an effective parameter. Fully describing evanescent excitations is beyond the tunneling limit, but could be approximated by assuming that $k_s$ is modified to match the edge state momentum.

\section{Conclusion}
\label{sec:conclusions}

We have derived an effective many-body Hamiltonian for an integer QH edge state at filling factor $\nu=1$ in tunneling contact with an $s$-wave SC with SOC and Meissner current at the surface. We show that taking these two latter elements into account resolves the apparent contradiction between experimentally observed induced superconducting correlations from an $s$-wave SC on the one hand, and the theoretical conditions of spin- and momentum conservation in the tunneling limit on the other.

We have analytically predicted the probability of an injected electron to be converted into a hole while propagating along the proximitized edge state, showing that when the electron energy is near the middle of the SC gap, this process is suppressed by Pauli blocking. By applying a modified Landauer-B\"uttiker formalism, we have related this hole conversion probability to the experimentally accessible quantity of differential downstream resistance. A key prediction is that despite the low interface transparency implied by the tunneling limit, strong electron-hole conversion can be reached via scattering with the Andreev edge channel at a resonance condition determined by the Meissner current.

\section*{Acknowledgments}
The authors wish to thank Harold Baranger, Lingfei Zhao and Pablo Burset for fruitful discussions. ABM and TLS acknowledge support from the National Research Fund, Luxembourg under the grant ATTRACT , Grant No. A14/MS/7556175/MoMeSys. ABM and BB acknowledge support from St.~Leonard's European Inter-University Doctoral Scholarship of the University of St.~Andrews. PR acknowledges financial support by the Deutsche Forschungsgemeinschaft (DFG, German Research Foundation) within the framework of Germany’s Excellence Strategy-EXC-2123 QuantumFrontiers-390837967.

The work presented in this paper is theoretical. No data were produced, and supporting research data are not required.

\appendix

\section{The renormalized Numerov method}
\label{app:numerov}

We use the renormalized Numerov method \cite{johnsonNew1977} to integrate the Schr\"odinger equation (SE) by discretizing space into a grid defined by $x_n = x_0 + n h$, where $x_0$ is an initial point, $n \in \mathbb{Z}$ and $h$ is a small step, in this case chosen as $h = 10^{-4} \ell$. Taylor expanding the discrete wave function and dropping terms of order $\mathcal{O}(h^6)$ or higher, we use the SE to formulate an iterative algorithm where the ratio $R_n = \psi_{n+1}/\psi_n$ is fully determined by $R^{-1}_{n-1}$. For a hard wall at $x=0$, we know that for any edge state we have $\psi(x=0) = \psi(x=15 \ell) = 0$, where $x=15 \ell$ is an arbitrarily chosen point in the bulk. This allows us to iterate forwards from from the hard wall and backwards from the bulk, both to the point $x_c$ which is chosen as the first extremum from the wall for numerical stability.

At this point we can define a matching function
\begin{align}
G(E) = \left( \frac{\psi(x_c+h)}{\psi(x_c)} \right)_{\text{left}} - \left( \frac{\psi(x_c+h)}{\psi(x_c)} \right)_{\text{right}},
\end{align}
which compares the (discrete) slopes of the iterated wave function from the two directions. Since $G(E) = 0$ if and only if $E$ is an eigenvalue of the Schr\"odinger equation, we can find the roots of $G(E)$ to obtain the eigenvalues of the Hamiltonian, after which the eigenfunctions can be constructed from the boundary conditions, the ratios $R_n$ and normalization.

\section{Integrating over the Euclidian action}
\label{app:action}
We can describe the effect of the superconductor on the edge state by integrating out the superconductor fields of the total action of the system, resulting in an effective action for the QH edge state \cite{aliceaNew2012}.

We consider the following Hamiltonian written as a functional
\begin{align}
H^{(1)}[ \psi \psi^\dagger  d d^\dagger]
= H_\text{edge}[\psi\psi^\dagger ] + H_\text{sc}^{(1)}[ d d^\dagger] + H_\text{tunn}[ \psi \psi^\dagger  d d^\dagger],
\end{align}
where the first two terms are given by \cref{eq:H1D,eq:Hsurf1} respectively, and $H_\text{tunn}$ from \cref{eq:HGamma} has been rewritten in terms of the $ d_{\v{k},1/2}$ quasiparticle operator as defined by \cref{eq:quasiparticleBasis}. We can construct the Euclidian action corresponding to $H^{(1)}$ as \cite{altlandChapter2010}
\begin{multline}
S[\psi^\dagger \psi d^\dagger d] \\
= \sum_{\v{k},q,\omega,j} i \omega (\psi_{q,\omega}^\dagger \psi_{q,\omega} + d^\dagger_{\v{k},j,\omega} d_{\v{k},j,\omega}) - H^{(1)}[\psi \psi^\dagger d d^\dagger ],
\end{multline}
where $\omega$ is a Matsubara frequency. Note that in the action, $\psi$ and $d$ are considered as Grassmann variables instead of fermionic fields. To perform the path integral over $e^{-S}$ it is useful to re-express the action such that the integral becomes a Gaussian integral. We first write the action as $S[\psi^\dagger \psi d^\dagger d] = S_0[\psi^\dagger \psi] + \delta S[\psi^\dagger \psi d^\dagger d]$, where $S_0$ contains all terms which only depend on the QH electron fields.
We then consider the quantum partition function path integral
\begin{align}
\mathcal{Z} =  \int D[\psi \psi^\dagger d d^\dagger]\ e^{-S_0[\psi \psi^\dagger] - \delta S[\psi \psi^\dagger d d^\dagger ]},
\end{align}
where
\begin{align}
D[\psi \psi^\dagger d d^\dagger] = \prod_{\v{k},q,\omega} d\psi^\dagger_{q,\omega}\ d\psi_{q,\omega} \ d (d^\dagger_{\v{k},\omega})\ d (d_{\v{k},\omega}).
\end{align}
The effective action is obtained by performing the path integral over the superconductor fields
\begin{multline}
 \int D[d d^\dagger]\ e^{-\delta S[\psi \psi^\dagger d d^\dagger]}
\\
= \prod_{{\v{k}, \omega}} \int d (d_{\v{k},\omega}^\dagger) d (d_{\v{k},\omega}) e^{-\delta S[\psi \psi^\dagger d d^\dagger]}.
\end{multline}
To solve this integral, we first rewrite $\delta S$ in terms of the vectors
\begin{align}
\phi_{\v{k},\omega} &= \begin{pmatrix}
 d_{\v{k},1,\omega}\\
 d^\dagger_{-\v{k},2,-\omega}
\end{pmatrix},\\
\nu_{\vkappa, \omega}^\dagger &=
-\tilde \gamma \begin{pmatrix}
iu_\vkappa e^{-i \tilde\theta_+}  \psi^\dagger_{k_S+q,\omega}
-  v_\vkappa^* e^{i\tilde \theta_+}  \psi_{k_S-q,-\omega} \\
-iu^*_\vkappa e^{i\tilde \theta_-}  \psi_{k_S-q,-\omega}
+ v_\vkappa e^{-i\tilde \theta_-}  \psi^\dagger_{k_S+q,\omega}
\end{pmatrix}^T,
\end{align}
where we have defined the angle
\begin{align}
\tilde \theta_\pm = \frac{1}{2}\arg\left[k_s\pm(q+ik_z)\frac{\epsilon_{\vkappa}+E_s}{\zeta_{\vkappa}}\right],
\end{align}
and $\vkappa = (q,k_y,k_z)$. The appearance of $\vkappa$ instead of $\v{k}$ arises from the summation over the Kronecker delta in the tunneling Hamiltonian. Defining the diagonal matrix
\begin{align}
\Lambda_{\v{k},\omega} &= \begin{pmatrix}
-i \omega + \epsilon^{(1)}_{\v{k},\pm} & 0\\
0 & -i \omega - \epsilon^{(1)}_{\v{k},\pm}
\end{pmatrix},
\end{align}
where $\epsilon^{(1)}_{\v{k},\pm}$ is given in \cref{eq:SCspectrum}, allows us to write the effective action as
\begin{multline}
\delta S[\phi \phi^\dagger \nu \nu^\dagger] \\
= \sum_{\v{k},q, \omega}  \Big[ \phi_{\v{k},\omega}^\dagger \Lambda_{\v{k},\omega} \phi_{\v{k},\omega} - \phi_{\vkappa,\omega}^\dagger \nu_{\vkappa,\omega} - \nu_{\vkappa, \omega}^\dagger \phi_{\vkappa,\omega} \Big].
\end{multline}
Since the order of multiplication in the path integral measure is irrelevant, we can first take the product of all momenta with $k_x = q$, written as $\vkappa$, and afterwards multiply by the remaining momenta where $k_x \neq q$, written as $\v{k}^\prime$. This can be explicitly written as
\begin{align}
\prod_\v{k} = \prod_\vkappa \prod_\v{k^\prime},
\end{align}
which lets us split the product of integrals as follows
\begin{multline}
\int D[\phi^\dagger \phi] e^{-\delta S[\phi \phi^\dagger \nu \nu^\dagger]}
\\
= \prod_{{\vkappa, \omega}} \int d \phi_{\vkappa,\omega}^\dagger d\phi_{\vkappa,\omega} e^{\delta S_1 [\phi \phi^\dagger \nu \nu^\dagger]}
\\
\times
\prod_{\v{k}^\prime,\omega} \int d \phi_{\v{k}^\prime,\omega}^\dagger d\phi_{\v{k}^\prime,\omega} e^{\delta S_2 [\phi \phi^\dagger \nu \nu^\dagger]},
\end{multline}
where
\begin{align}
\delta S_1 &= -\phi_{\vkappa,\omega}^\dagger \Lambda_{\vkappa,\omega} \phi_{\vkappa,\omega} + \phi_{\vkappa,\omega}^\dagger \nu_{\vkappa, \omega} + \nu_{\vkappa,\omega}^\dagger \phi_{\vkappa,\omega} , \\
\delta S_2 &= -\phi_{\v{k}^\prime,\omega}^\dagger \Lambda_{\v{k}^\prime,\omega} \phi_{\v{k}^\prime,\omega}.
\end{align}
This reduces the integration to a Gaussian integral of the type
\begin{multline}
\int D[\phi^\dagger \phi] e^{-\delta S[\phi \phi^\dagger \nu \nu^\dagger]}
\\
= \exp \Big( \sum_{\v{k}^\prime, \vkappa, \omega} \Big[ \nu_{\vkappa, \omega}^\dagger \Lambda^{-1}_{\vkappa,\omega} \nu_{\vkappa, \omega} + C \Big] \Big),
\end{multline}
where
\begin{align}
C = \ln \Big(\det(\Lambda_{\vkappa,\omega}) \Big) + \ln\Big(\det(\Lambda_{\v{k^\prime},\omega})\Big).
\end{align}
Since $C$ contributes an overall, constant shift to the action (and thus the Hamiltonian), it can be ignored, leading to the partition function
\begin{align}
\mathcal{Z} = \int D[\psi^\dagger \psi]\ \exp\left(-S_0 - \sum_{\vkappa, \omega} \nu_{\vkappa,\omega}^\dagger \Lambda_{\vkappa,\omega}^{-1} \nu_{\vkappa,\omega} \right).
\end{align}
For a small tunneling amplitude $\Gamma$, the $\omega$-dependence in $\Lambda$ can be neglected, and we can recover an effective Hamiltonian from the effective action by considering the $\omega=0$ case. This amounts to neglecting retardation, and translates to assuming that the induced pairing velocity $\DeltaP$ as defined in \cref{eq:effDelta} is small compared with the edge state Fermi velocity $v_F$.

\section{Landauer-B\"uttiker formula with particles and holes}
\label{app:LBextended}
Calculating the current through a lead with both particles and holes is a well-known problem, already treated in the seminal paper by Blonder, Tinkham and Klapwijk \cite{blonderTransition1982}, and so we only touch upon a few details in this appendix for the sake of completeness. We specifically consider a chiral system, which means there is no reflection, and all incoming states are transmitted.

If we describe incoming particle-hole states with energy $E$ by the field $ i_{p/h,E}$ and outgoing states by the field $o_{p/h,E}$, the current expectation value can be constructed by counting incoming holes and outgoing particles as contributing negatively to the current,
\begin{multline}
\braket{I(t)} = \frac{1}{2 \pi} \int_0^\infty dE dE^\prime e^{i(E-E^\prime)t}\Big(
\braket{ i^\dagger_{p,E}  i_{p,E^\prime}}
\\- \braket{ o^\dagger_{p,E}  o_{p,E^\prime}}
- \braket{ i^\dagger_{h,E}  i_{h,E^\prime}}
+ \braket{ o^\dagger_{h,E}  o_{h,E^\prime}}
\Big).
\end{multline}
The incoming particles and holes obey the distributions
\begin{align}
\braket{ i^\dagger_{p,E}  i_{p,E^\prime}} &= \delta(E-E^\prime) n_F(E-\mu), \\
\braket{ i^\dagger_{h,E}  i_{h,E^\prime}} &= \delta(E-E^\prime) n_F(E+\mu).
\end{align}
These relations can be extended straightforwardly to $N$ leads with respective chemical potentials $\{\mu_1, \ldots, \mu_N\}$. The relation between incoming states in any lead and outgoing states in lead $m$ are then given by the scattering matrix
\begin{align}
\begin{pmatrix}
o_{m,p,E}\\
o_{m,h,E}
\end{pmatrix}
= \sum_n \begin{pmatrix}
t_{mn,pp} & t_{mn,ph}\\
t_{mn,hp} & t_{mn,hh}
\end{pmatrix}
\begin{pmatrix}
i_{n,p,E}\\
i_{n,h,E}
\end{pmatrix}.
\end{align}
From this one can show that the steady-state current in lead $m$ becomes
\begin{align}\label{eq:LB_Im}
\braket{I_m(t)} &= \frac{e}{h} \sum_n \int_0^\infty dE \Big[ n_F(E-\mu_m)  - n_F(E+\mu_m) \notag \\
&- (|t_{mn,pp}|^2-|t_{mn,hp}|^2) n_F(E-\mu_n) \notag \\
&-(|t_{mn,ph}|^2-|t_{mn,hh}|^2)n_F(E+\mu_n) \Big],
\end{align}
which generalizes the Landauer-B\"uttiker formula to superconductors described in the Nambu basis.

\bibliography{chiral-andreev-paper-long}

\end{document}